\documentclass[10pt,conference]{IEEEtran}
\IEEEoverridecommandlockouts
\usepackage{cite}
\usepackage{comment}
\usepackage{amsmath,amssymb,amsfonts}
\usepackage{algorithm}
\usepackage{algpseudocode}
\usepackage{graphicx}
\usepackage{float}
\usepackage{textcomp}
\usepackage{xcolor}
\usepackage{tikz}
\usetikzlibrary{arrows.meta, positioning, fit, calc, backgrounds}
\usepackage{siunitx}
\usepackage{caption}
\usepackage[nolist]{acronym}
\usepackage{url}
\usepackage{subcaption}
\usepackage{booktabs}

\renewcommand{\thetable}{\arabic{table}}

\usepackage[letterpaper,top=0.75in,bottom=1.04in,left=0.625in,right=0.625in,columnsep=0.25in]{geometry}

\def\BibTeX{{\rm B\kern-.05em{\sc i\kern-.025em b}\kern-.08em
    T\kern-.1667em\lower.7ex\hbox{E}\kern-.125emX}}

\newcommand\copyrightnotice{%
\begin{tikzpicture}[remember picture,overlay]
\node[anchor=south,yshift=10pt] at (current page.south) {%
\footnotesize \parbox{0.9\textwidth}{\centering
Accepted to be published in: 2026 IEEE 37th International Symposium on Personal, Indoor and Mobile Radio Communications (PIMRC), Singapore, Sept. 1--4, 2026. \\
\textcopyright~2026 IEEE. Personal use of this material is permitted.
Permission from IEEE must be obtained for all other uses, in any current or
future media, including reprinting/republishing this material for advertising
or promotional purposes, creating new collective works, for resale or
redistribution to servers or lists, or reuse of any copyrighted component
of this work in other works.}};
\end{tikzpicture}}

\begin{document}

\bstctlcite{IEEEexample:BSTcontrol}

\title{Optimal Measurement Point Selection for Radio Environment Maps: A Two-Stage Approach}
\vspace{-1em}
\author{\IEEEauthorblockN{Bastian Perner, Friedemann Laue, Maximilian Lübke, and Norman Franchi}\\
\vspace{-1em}
\IEEEauthorblockA{\textit{Institute for Smart Electronics and Systems}\\
\textit{Friedrich-Alexander-Universität Erlangen-Nürnberg}\\
Email: \{bastian.perner, friedemann.laue, maximilian.luebke, norman.franchi\}@fau.de}
}

\maketitle
\copyrightnotice
\vspace*{-\baselineskip}
\pagestyle{empty}
\thispagestyle{empty}

\begin{acronym}

\acro{CR}{Cognitive Radio}
\acro{PPDR}{Public Protection Disaster Relief}
\acro{EM}{Electromagnetic}
\acro{REM}{Radio Environment Map}
\acro{IDW}{Inverse Distance Weighting}
\acro{ILP}{Integer Linear Program}
\acro{5G}{fifth generation mobile network}
\acro{6G}{sixth generation mobile network}
\acro{NPN}{non-public network}
\acro{OED}{Optimal Experimental Design}
\acro{ECS}{Emergency Communication System}
\acro{BNetzA}{Bundesnetzagentur}
\acro{PSD}{Power Spectral Density}
\acro{SSB}{Synchronization Signal Block}
\acro{MAE}{Mean Absolute Error}
\acro{WSN}{Wireless Sensor Networks}
\acro{NN}{Nearest Neighbor Interpolation}
\acro{SS-RSRP}{Synchronization Signal Received Signal Reference Power}
\acro{GNSS}{Global Navigation Satellite System}

\end{acronym}

\begin{abstract}
    \ac{EM} situational awareness is essential for the deployment and operation of local and temporary \acp{NPN}. \acp{REM} provide an effective basis for network planning and operation by capturing the prevailing \ac{EM} conditions. However, efficient measurement location selection for \ac{REM} construction remains a key challenge, particularly under strict time and cost constraints. This paper presents a two-stage measurement selection framework that employs \ac{OED}-based strategies to generate a reduced candidate set during a pre-selection phase. In a subsequent optimization stage, combinatorial optimization is applied to select an improved subset of measurement locations by minimizing the \ac{MAE} of the resulting \ac{REM}, using a scenario-specific propagation simulation model. The results demonstrate that the proposed framework reduces the number of required measurement locations while consistently improving \ac{REM} construction accuracy across all evaluated strategies. Validation using real-world measurement data confirms the practical applicability of the approach, with the optimization stage achieving \ac{MAE} reductions from about 4\% to around 50\%, independent of the chosen pre-selection strategy.

\end{abstract}

\acresetall

\begin{IEEEkeywords}
    5G, Cooperative Spectrum Sensing, Measurement Planning, Non-Public Networks, Radio Environment Map, Sensor Placement, Simulation-Based Optimization, Spectrum Cartography
\end{IEEEkeywords}

\section{Introduction}
    \label{sec:introduction}
    One important use case of \ac{5G} and \ac{6G} communication standards is \acf{NPN}, particularly for local and temporally limited deployments. Such \acp{NPN} require \acf{EM} situational awareness to ensure reliable spectrum usage, coexistence with other \acp{NPN} and public networks, and short-term network planning~\cite{shatov_integrated_2023,9604989}. Typical application scenarios include private deployments, industrial testbeds, and event-driven setups~\cite{horbach_opportunistic_2024,wei2025survey}. Event-driven installations, as described in~\cite{horbach_opportunistic_2024} in the context of \acf{ECS}, impose stringent constraints on time, measurement effort, and prior environmental knowledge. To establish \ac{EM} situational awareness, \acfp{REM} are commonly employed. These multi-dimensional representations capture spatial, spectral, and temporal characteristics of the \ac{EM} environment.

\acp{REM} enable the assessment of white spaces and interference, thereby supporting network planning, coexistence analysis, and spectrum monitoring~\cite{wei2025survey}. Accurate \ac{REM} construction requires spatially distributed measurements, but exhaustive measurement campaigns are often impractical due to time and cost constraints. Consequently, the selection of measurement locations constitutes a fundamental challenge in \ac{REM} construction. Prior work has addressed this challenge through heuristic, sampling-based selection strategies~\cite{suchanskiRadioEnvironmentMaps2020,Shen2022SamplingOpt} as well as minimization approaches on fixed spatial meshes~\cite{MARTINEZGONZALEZ2022113577}. In our previous work, we introduced two \acf{OED}-based, environment-agnostic selection strategies focusing exclusively on coverage- and sparsity-aware criteria~\cite{perner_measplanning_2025}.

While this environment-agnostic approach requires no additional information and is applicable across different scenarios, it neglects scenario-specific propagation characteristics. In contrast, incorporating model-based information has been shown to improve \ac{REM} construction quality. Examples include sparsity-aware Bayesian models~\cite{Wang2024Sparse3DREM}. Propagation simulation tools can exploit available environmental information such as building layouts, digital elevation models, or antenna configurations to approximate the \ac{EM} situation~\cite{bektas2021}.

Direct optimization over the full set of accessible measurement locations $P$ is computationally intractable: for $|P| = 200$ and a target subset size $k = 20$, the number of candidate subsets $\binom{|P|}{k}$ exceeds $10^{25}$. This motivates a two-stage approach that first reduces the search space through environment-agnostic heuristics, enabling tractable optimization in a second stage.

This work combines both approaches by repurposing environment-agnostic selection strategies to generate a reduced candidate set of measurement locations. In a second stage, combinatorial optimization refines this candidate set using environmental characteristics derived from propagation simulation. This split enables computationally feasible optimization while preserving the benefits of heuristic selection. The simulation does not replace field measurements but solely guides the selection of measurement locations, while the final \ac{REM} is always constructed from actual measurements.

Specifically, this work makes the following contributions:
\begin{itemize}
\item Enhancement of existing selection strategies through mitigation of initialization effects.
\item A two-stage measurement selection framework combining strategy-based candidate pre-selection with optimization over the resulting candidate set.
\item Integration of simulation-derived environmental information into the optimization procedure.
\item Validation on real-world measurement data, confirming the transferability of the simulation-based selection to practical deployments.
\end{itemize}

The remainder of this paper is structured as follows. Section~\ref{sec:concept} introduces the problem formulation and enhanced selection strategies. Section~\ref{sec:measurements} describes the evaluation setup, followed by the results in Section~\ref{sec:evaluation}. Section~\ref{sec:discussion} discusses the findings, and Section~\ref{sec:conclusion} concludes the paper with directions for future work.

\section{Problem Formulation and proposed Method}
    \label{sec:concept}
    The proposed two-stage approach consists of a pre-selection phase followed by an optimization phase. In the pre-selection phase, dedicated strategies select a subset of $N$ candidate measurement locations from the full set $P$ based solely on spatial distribution criteria, requiring no propagation knowledge.

In the optimization phase, propagation information is incorporated via propagation simulation to perform combinatorial optimization over the pre-selected candidates, yielding an optimized subset of size $k$ from the candidate set of size $N$. The simulation incorporates environment geometry and antenna radiation patterns. For the rural open-field scenario considered here, the environment model captures terrain elevation profiles. Urban or indoor scenarios would additionally incorporate building layouts and material properties. Antenna models reflect the deployed hardware specifications.

\subsection{Problem Definition}
Let $\hat{A}$ denote the continuous target monitoring area. To construct a \ac{REM}, $\hat{A}$ is discretized into a set $A$ of regularly spaced grid points representing the evaluation locations. Furthermore, $P \subset \hat{A}$ denotes the set of accessible measurement locations, and $S \subset P$ denotes a general subset of measurement locations. For a fixed measurement budget $k \in \mathbb{N}$, we define $S_k \subset P$ with $|S_k| = k$ as a size-$k$ selection of measurement locations. The \ac{REM} constructed from a subset $S$, denoted $\text{REM}_S$, is obtained by \ac{IDW} interpolation, chosen for its low computational complexity, suitability for irregularly placed sampling points, and independence from statistical assumptions about the underlying field. The framework is designed to be interpolation-agnostic, as the optimization does not rely on the internal structure of the interpolation operator. Empirical validation with alternative interpolation methods is left for future work.

Reconstruction quality is evaluated on the grid $A$ using the \acf{MAE} between the constructed \ac{REM} and a reference \ac{REM} $\text{REM}_\text{ref}$:
\begin{equation} \label{eq:mae}
    \mathrm{MAE}(S) =
    \frac{1}{|A|}
    \sum_{a \in A}
    \big|\, \mathrm{REM}_\text{ref}(a) - \mathrm{REM}_S(a) \,\big|.
\end{equation}
For the synthetic dataset, $\text{REM}_\text{ref}$ is the simulated field. For the real-world dataset, no full-field ground truth is available, so $\text{REM}_\text{ref}$ is obtained by interpolating all accessible measurements. The resulting partial self-reference is discussed as a limitation in Section~\ref{sec:discussion}.

The objective is to identify the optimal subset $S_k^* \subset P$ of $k$ measurement positions that minimizes $\mathrm{MAE}(S_k)$. As motivated in Section~\ref{sec:introduction}, direct optimization over $P$ is computationally intractable for typical values of $|P|$ and $k$.

\subsection{Heuristic Restriction of the Search Space (Stage~1)}
The solution space is restricted to a candidate subset prior to optimization. Heuristic selection strategies based on spatial distribution criteria generate a reduced candidate set:
\begin{equation}
   P_N \subset P, \quad N \ll |P|.
\end{equation}

Two classes of candidate selection strategies from~\cite{perner_measplanning_2025} are employed: greedy baselines based on \ac{OED} criteria, and enhanced variants with iterative convergence refinement (detailed below). Given a partially constructed set $S_i \subset P$ with $|S_i| = i$ and a distance metric $d(\cdot, \cdot)$, the next measurement location $s_{i+1}$ is selected as follows.

The \textit{MaxMin} strategy maximizes the minimum distance to all previously selected locations, thereby promoting spatial spread:
\begin{equation}
    s_{i+1} = \arg\max_{p \in P \setminus S_i} \min_{s \in S_i} d(p, s).
\end{equation}

The \textit{MinMax} strategy minimizes the maximum distance to the nearest selected location, thereby reducing the largest coverage gap:
\begin{equation}
    s_{i+1} = \arg\min_{p \in P \setminus S_i} \max_{s \in S_i} d(p, s).
\end{equation}

Both strategies are initialized with a randomly selected first location $s_1 \in P$, which introduces a dependency on this initial choice.

The enhanced variants (\textit{MaxMinConv}, \textit{MinMaxConv}) mitigate this initialization dependency through the iterative refinement summarized in Algorithm~\ref{alg:refinement}, with $f$ the respective \ac{OED} score. Starting from the greedy baseline $P_N^{(0)}$, each candidate is swapped for a better replacement until no single-point swap improves $f$. The procedure is guaranteed to converge in a finite number of iterations.

\begin{algorithm}[t]
\caption{Refinement for candidate set generation.}
\label{alg:refinement}
\begin{algorithmic}[1]
\State \textbf{Input:} Baseline output $P_N^{(0)}$, criterion $f$
\Repeat
    \State $\textit{improved} \gets \text{false}$
    \ForAll{$p \in P_N^{(t)}$}
        \State $P_{\text{temp}} \gets P_N^{(t)} \setminus \{p\}$
        \State $p' \gets \arg\max_{q \in P \setminus P_{\text{temp}}} f(P_{\text{temp}} \cup \{q\})$
        \If{$f(P_{\text{temp}} \cup \{p'\}) > f(P_N^{(t)})$}
            \State $P_N^{(t+1)} \gets P_{\text{temp}} \cup \{p'\}$
            \State $\textit{improved} \gets \text{true}$
        \EndIf
    \EndFor
\Until{not \textit{improved}}
\State \textbf{Output:} Refined candidate set $P_N$
\end{algorithmic}
\end{algorithm}

\subsection{Optimization over Candidate Measurement Sets (Stage~2)}
The second stage selects a size-$k$ subset from the candidate set $P_N$ generated in Stage~1. The target is the minimization of the reconstruction error,
\begin{equation} \label{eq:objective}
    S_k^* = \arg\min_{S \subseteq P_N,\, |S| = k} \mathrm{MAE}(S),
\end{equation}
which is approximated heuristically, as the \ac{IDW}-based $\mathrm{MAE}$ is non-convex in the discrete selection $S$ and cannot be cast as a linear objective. Stage~2 therefore proceeds in two steps. First, an \ac{ILP} enforces the cardinality constraint $|S| = k$ over the $N$ candidates and returns a feasible size-$k$ selection. A local search then refines this selection by iteratively exchanging a selected for an unselected candidate whenever this reduces $\mathrm{MAE}$. The procedure yields a local optimum of Eq.~\eqref{eq:objective} rather than a global optimum, and is performed independently for each considered subset size $k \leq N$.

\subsection{Computational Complexity}
The computational cost of each stage is as follows. In Stage~1, the greedy baseline requires $\mathcal{O}(N \cdot |P|)$ distance evaluations. The convergence refinement adds $\mathcal{O}(I \cdot N \cdot |P|)$ evaluations, where $I$ denotes the number of refinement iterations until convergence.

In Stage~2, the two-step heuristic of the preceding subsection avoids combinatorial enumeration. For each $(N,k)$, the \ac{ILP} solve is followed by at most $R$ local-search passes, each evaluating $\mathcal{O}(k\,(N-k))$ candidate swaps at a cost of $\mathcal{O}(|A|\,k)$ per \ac{MAE} evaluation. The full subset-size sweep over $k \in [1,N]$ is therefore polynomial in $N$, $|A|$, and $R$ and, unlike an exhaustive search, remains tractable across the entire sweep. In contrast, direct optimization over the full set $P$ would require evaluating $\binom{|P|}{k}$ subsets, which is intractable for typical values of $|P|$.

\section{Evaluation Setup}
    \label{sec:measurements}
    This section describes the evaluation setup. The scenario and database are based on the real-world measurement campaign presented in our previous work~\cite{perner_measplanning_2025, horbach_opportunistic_2024}. A simulation model replicates this campaign to generate synthetic propagation data. Measurement locations are fixed across both simulation and real-world datasets.

\subsection{Database}
The database consists of real-world measurement data and synthetic simulation data derived from a model of the same measurement campaign. Both datasets share identical spatial discretization and measurement locations.

\subsubsection{Measurement Campaign Data}
\begin{figure}[t]
    \centering
    \includegraphics[width=0.4\textwidth]{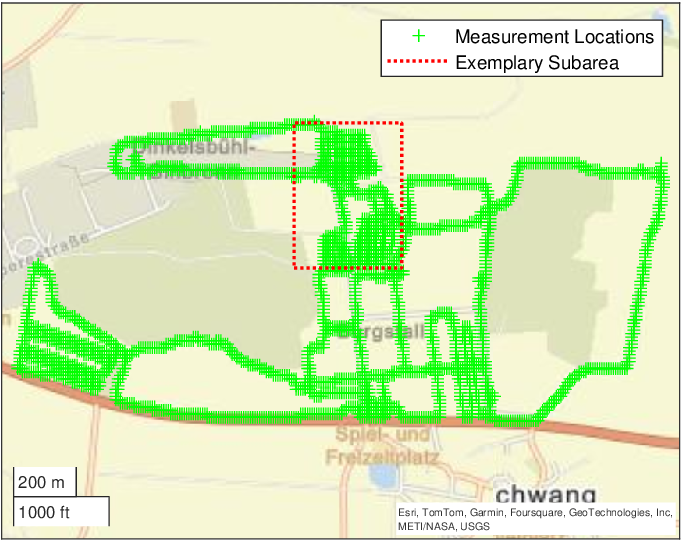}
    \caption{Overview of available measurement locations and the exemplary stage area used in the evaluation.}
    \label{fig:trajectory}
\end{figure}

Real-world data were collected during a measurement campaign at a music festival near Dinkelsbühl (Bavaria, Germany), representing a rural scenario (Fig.~\ref{fig:trajectory}). Measurements are associated with coordinates obtained from \ac{GNSS} position estimates. The dataset comprises approximately $|P| \approx 2{,}400$ \ac{SS-RSRP} measurement locations, obtained by spatially aggregating raw \ac{GNSS} positions to a resolution of approximately \SI{11}{\meter} (four decimal places in latitude and longitude).

\subsubsection{Synthetic Propagation Data}
Synthetic data were generated using WinProp (Altair Feko Suite 2025). The simulation incorporates:
\begin{itemize}\setlength{\itemsep}{0pt}\setlength{\parskip}{0pt}
    \item Environment model based on terrain elevation data from Bayern Atlas\footnote{\url{https://geoportal.bayern.de/bayernatlas}}.
    \item Antenna model reflecting the deployed hardware characteristics.
    \item Dominant path propagation model.
\end{itemize}
Simulated received signal strength values were computed at the same measurement locations as in the real-world campaign.

\subsection{Evaluation Metrics and Procedure}
Evaluation is based on the \ac{MAE} defined in Eq.~\eqref{eq:mae}, computed for both the synthetic and the real-world dataset. Results are averaged across 17 areas within the festival grounds, including the stage area, camping sites, and the full area. In addition, the number of selected measurement locations is used to compare strategy-based and optimization-based approaches.

Finally, \ac{MAE} differences between pre-selection (Stage~1) and simulation-based optimization (Stage~2) are evaluated on real-world data to assess the practical applicability of the proposed two-stage framework.

\section{Evaluation}
    \label{sec:evaluation}
    This section presents the results of the pre-selection strategies (Stage~1), the optimization (Stage~2) with respect to the size of the candidate set and final subset, and the applicability of simulation-based optimization to real-world data.

\subsection{Strategy-Based Pre-selection Performance (Stage~1)}
\begin{figure}
    \centering
    \includegraphics[width=0.875\linewidth]{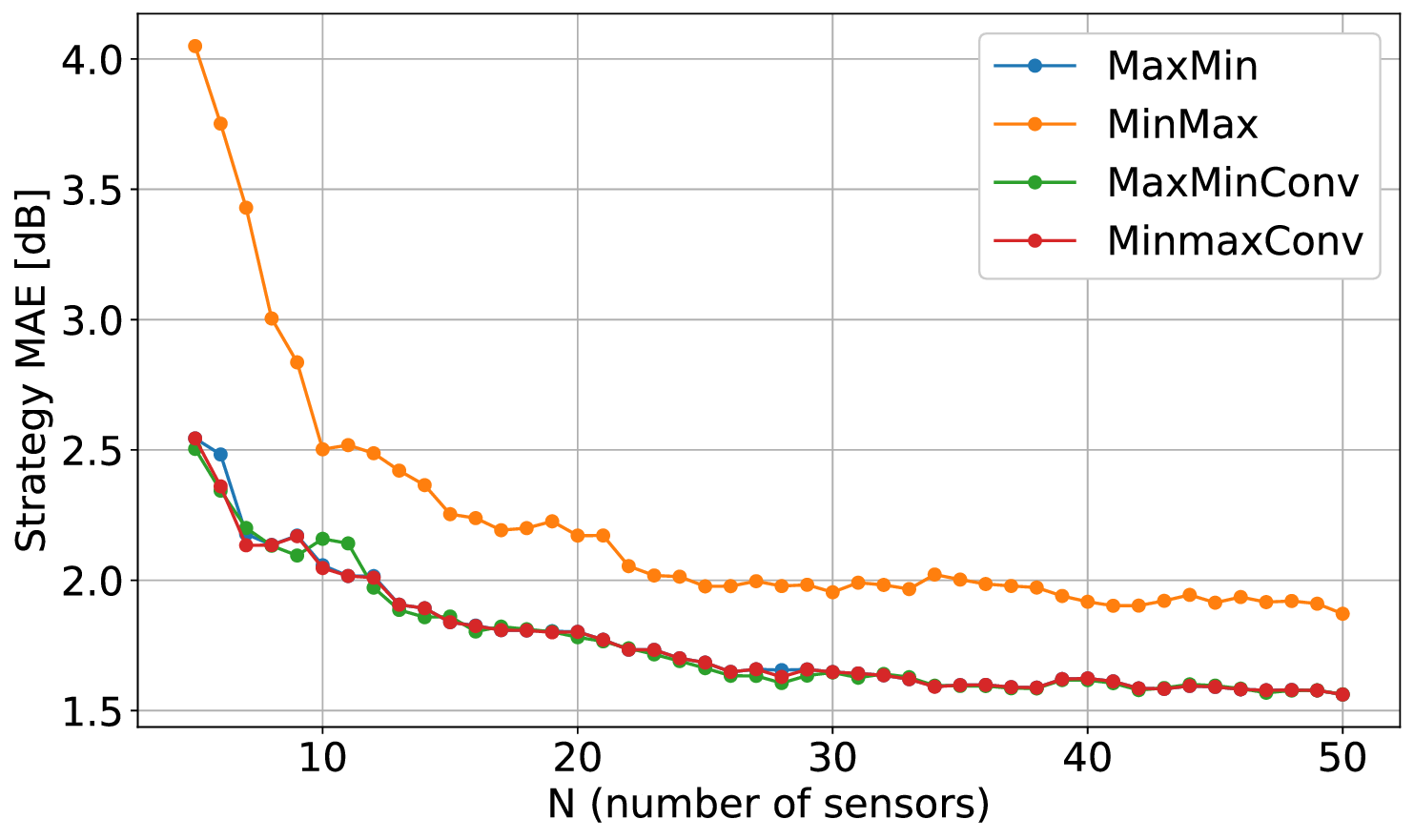}
    \caption{Comparison of \ac{REM} construction \ac{MAE} across number of measurement locations $N$ for legacy and enhanced strategies.}
    \label{fig:strategies}
\end{figure}

The two legacy pre-selection strategies (\textit{MaxMin}, \textit{MinMax}) and the proposed enhanced variants (\textit{MaxMinConv}, \textit{MinMaxConv}) are evaluated by comparing the \ac{MAE} of the \ac{REM} constructions across different numbers of measurement locations $N \in [5,50]$. At this stage, no optimization is applied. The evaluation solely reflects the effect of the pre-selection strategies themselves. Results are shown in Fig.~\ref{fig:strategies}.

While \textit{MaxMin}, \textit{MinMaxConv} and \textit{MaxMinConv} strategies show a \ac{MAE} in the range of \SIrange{1.5}{2.6}{\dB}, the \ac{MAE} of \textit{MinMax} lies between \SIrange{1.8}{4.1}{\dB}. There is no measurable difference in \ac{MAE} between the \textit{MaxMin} strategy and its enhanced variant \textit{MaxMinConv}. In contrast, enhancing the \textit{MinMax} strategy yields a clear performance improvement, bringing its accuracy to a level comparable with \textit{MaxMin} and \textit{MaxMinConv}.

The results indicate that the proposed enhancement depends strongly on the original selection principle. For \textit{MaxMin}, the convergence step does not yield measurable improvement. The \textit{MinMax} strategy, however, benefits substantially from the convergence mechanism, as it removes the dependency on the initial measurement location, which appears to have a higher impact on this strategy. This mitigates a structural weakness of the legacy \textit{MinMax} approach.

This gap stems from the strategies' spatial principles: \textit{MaxMin} maximizes the minimum inter-point distance, yielding uniform, interpolation-friendly distributions largely independent of initialization, whereas \textit{MinMax} fills the largest coverage gap and can cluster points depending on the initial location. The convergence step redistributes these clusters, elevating \textit{MinMaxConv} to a level comparable with \textit{MaxMin}. A comparison with random baseline selection in~\cite{perner_measplanning_2025} confirms that all evaluated strategies consistently outperform uninformed random selection.

\subsection{Optimization Analysis (Stage~2)}
After completing Stage~1, the optimization phase is applied to the candidate set $P_N$ with $N \in \{5,10,15,\ldots,50\}$ measurement locations, producing an optimized subset $S^*_k$ of size $k \in [5, N]$. To assess the optimization impact, the \textit{MinMax} and \textit{MaxMinConv} strategies are selected: \textit{MaxMinConv} exhibits representative behavior among the pre-selection strategies, while \textit{MinMax} shows slightly different characteristics. The optimization stage is evaluated on a single scenario (the entire festival area) with a fixed initial measurement location for all pre-selection strategies across $N$. Since the selection strategies are deterministic, $S_{i,\text{MinMax}} \subset S_{j,\text{MinMax}}$ holds for $i < j$.

\subsubsection{Joint Influence of Pre-selection and Final Subset Size}
\begin{figure}
    \centering
    \begin{subfigure}[b]{0.49\linewidth}
        \centering
        \includegraphics[width=\linewidth]{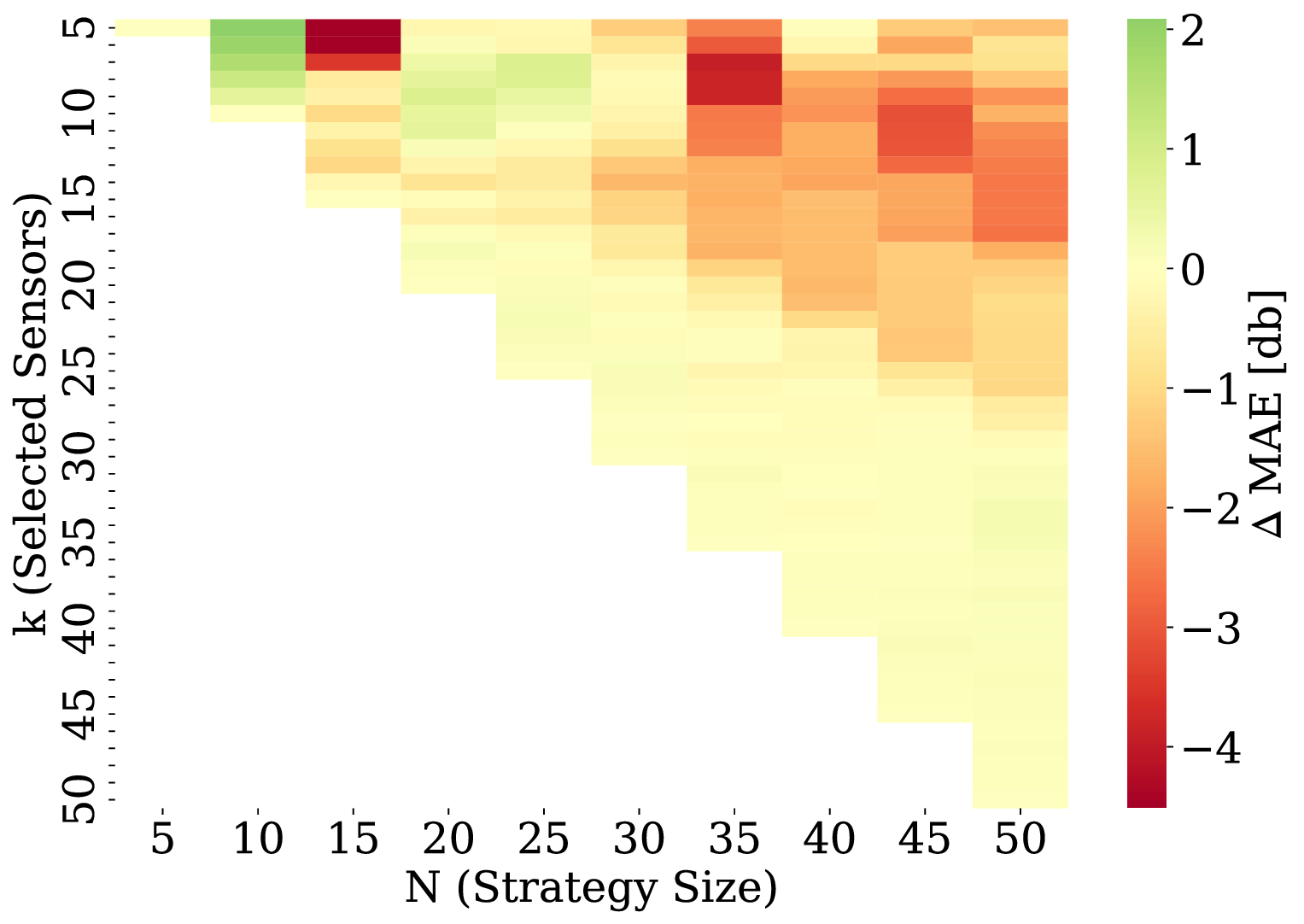}
        \caption{\textit{MinMax} strategy}
        \label{fig:heatmap_filter_7}
    \end{subfigure}
    \hfill
    \begin{subfigure}[b]{0.49\linewidth}
        \centering
        \includegraphics[width=\linewidth]{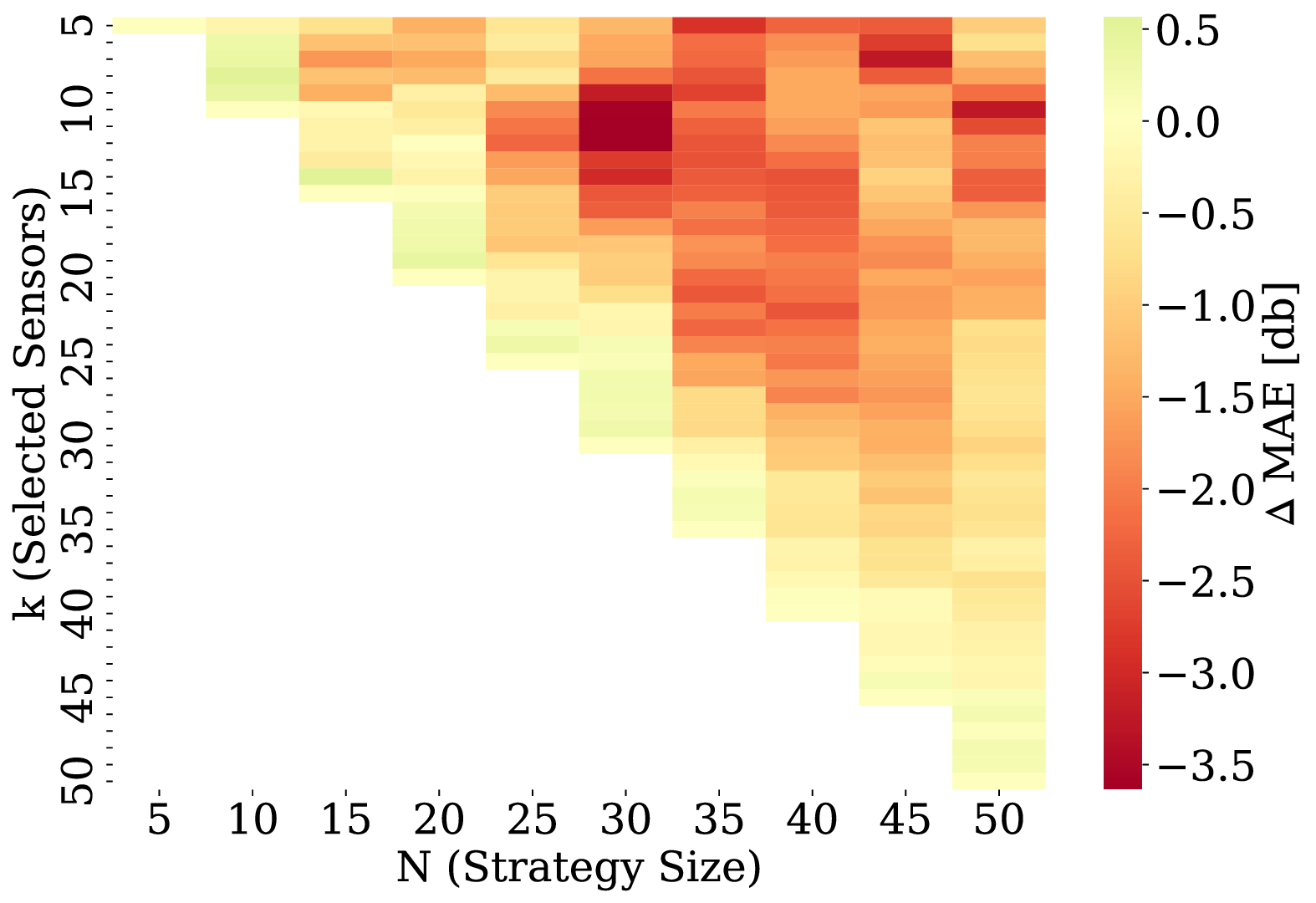}
        \caption{\textit{MaxMinConv} strategy}
        \label{fig:heatmap_filter_8}
    \end{subfigure}
    \caption{Heatmaps of the \ac{MAE} difference comparing optimization ($k$ measurements) to the pre-selection strategy ($N$ measurements) across different combinations of $N$ and $k$.}
    \label{fig:heatmap_comparison}
\end{figure}

Fig.~\ref{fig:heatmap_comparison} illustrates the change in \ac{MAE} for \ac{REM} construction achieved by the optimization phase (subset size $k$) relative to the pre-selection strategy alone (subset size $N$).

When $N=k$, both stages produce identical performance, as the optimization can only select $k=N$ locations from $N$ candidates. The heatmaps show that performance improvements generally increase with both $N$ and $k$. The largest degradations occur in the top-right region (high $N$, low $k$), while the greatest improvements appear when $k$ is slightly smaller than $N$. In certain $(N,k)$ combinations, the optimized subset with $k \leq N$ measurement locations outperforms the pre-selection subset with $N$ measurements.

For the \ac{IDW} interpolation used in this evaluation, the optimal subset size $k^*$ is consistently slightly below $N$. Since \ac{IDW} weights points inversely with distance, closely spaced pre-selected points dominate locally while adding little elsewhere, so removing such redundant points improves reconstruction. This redundancy is scarce at small $N$, where few candidates are available and each is essential, and at large $N$, where coverage is already dense, so $k^* \approx N$ in both regimes. The gap is largest at intermediate $N$, where the optimizer can exploit redundant candidates.

\subsubsection{Optimal Subset Size Selection}
\begin{figure}
\centering
\includegraphics[width=0.9\linewidth]{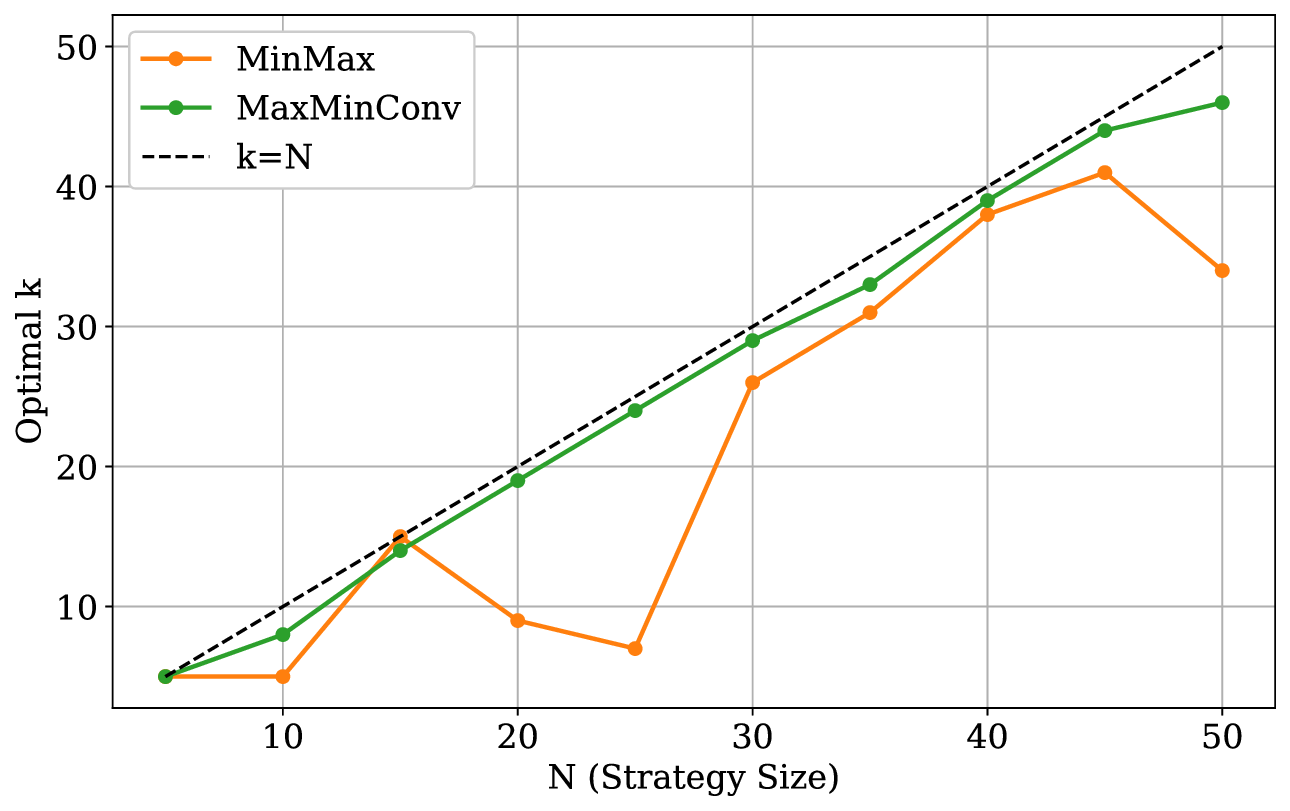}
\caption{Optimal $(N,k)$ combinations for the optimization phase using the \textit{MinMax} and \textit{MaxMinConv} pre-selection strategies.}
\label{fig:opt_k}
\end{figure}

While the joint influence of $N$ and $k$ was analyzed in the previous subsection, Fig.~\ref{fig:opt_k} shows the optimal $(N,k)$ combinations for each fixed value of $N$. The optimal $k$ stays close to $N$ for small and large $N$, whereas \textit{MinMax} exhibits a pronounced drop at intermediate $N \in [20,30]$, where a smaller subset $k^*$ already improves performance.

Across the evaluated range, the average optimal ratio $k/N$ is $0.94$ for \textit{MaxMinConv} (consistently between $0.90$ and $0.96$), whereas for \textit{MinMax} it averages $0.75$ and varies more strongly (between $0.57$ and $0.92$), dropping for intermediate $N$, which confirms the behavior observed in Fig.~\ref{fig:opt_k}.

\begin{figure}
\centering
\includegraphics[width=0.9\linewidth]{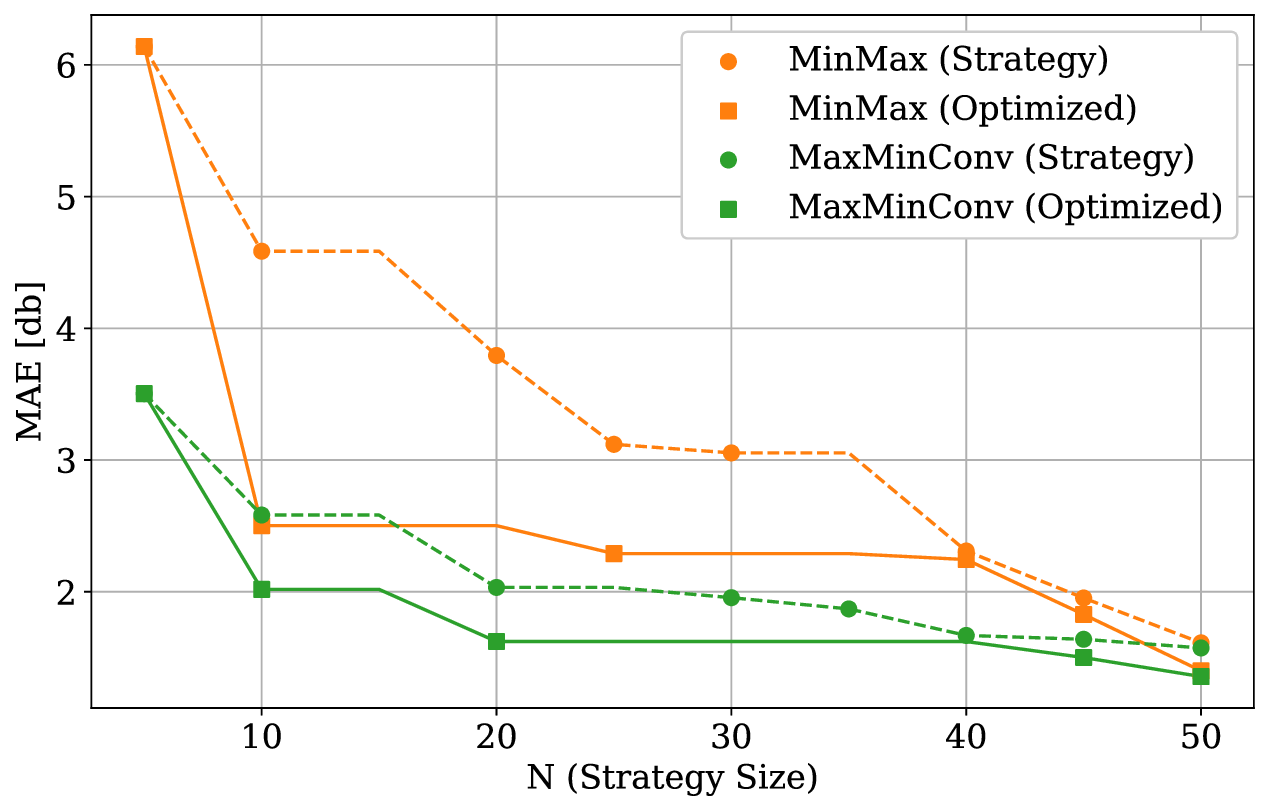}
\caption{Minimum \ac{MAE} of \ac{REM} construction achieved for the pre-selection strategies and the optimal $(N,k^*)$ pairs across all $N$ after the optimization phase for \textit{MinMax} and \textit{MaxMinConv} strategies.}
\label{fig:opt_N}
\end{figure}

Fig.~\ref{fig:opt_N} complements this analysis by showing the resulting \ac{REM} construction \ac{MAE} when using the optimized subset for each optimal $(N,k^*)$ pair and comparing it to the strategy-based \ac{REM} construction \ac{MAE}. For small values of $k$, increasing the subset size leads to substantial \ac{MAE} reduction, whereas marginal gains diminish for larger $k$. Saturation effects occur later for the \textit{MinMax} strategy compared to \textit{MaxMinConv}, both with and without optimization. After the optimization phase, the \ac{MAE} difference between \textit{MinMax} and \textit{MaxMinConv} is less than 0.5~dB, except for the boundary case of $N=5$.

\subsection{Simulation-to-Real Transferability}
The preceding evaluation was based exclusively on simulation data. The following results, derived from real-world measurement data, assess the transferability of the proposed approach to practical applications.

For a fixed pre-selection size $N$, the optimization procedure is applied using the \textit{MinMax} and \textit{MaxMinConv} pre-selection strategies. For each value of $N$, the optimal subset size $k$ is determined by minimizing the \ac{REM} construction \ac{MAE}, resulting in an optimized subset $S^*_k$ for the given pre-selection size. The analysis considers $N \in \{10,20,30,40,50\}$.

The resulting \ac{REM} construction \ac{MAE} is computed for the real-world measurement data using identical measurement locations after Stage~1 (pre-selection) and after Stage~2 (optimization). Tab.~\ref{tab:transfer} reports the absolute and relative differences between these two stages. In all evaluated cases, the optimization stage leads to an improvement in \ac{MAE}.

For the \textit{MaxMinConv} strategy, the largest improvement is observed at $N=20$, where the error is reduced by more than 50\%. However, the relative improvement varies considerably across $N$, ranging from approximately 4\% to over 50\%, indicating a less stable pre-selection performance. In contrast, the \textit{MinMax} strategy exhibits a consistent error reduction across all values of $N$, with the highest absolute and relative improvement occurring at $N=30$.

\begin{table}[t]
\centering
\caption{Improvement of REM construction MAE through simulation-based optimization for \textit{MinMax} and \textit{MaxMinConv} strategies}
\label{tab:transfer}
\setlength{\tabcolsep}{3pt}
\begin{tabular}{c | c c c c | c c c c}
\hline
& \multicolumn{4}{c|}{\textbf{MinMax}} & \multicolumn{4}{c}{\textbf{MaxMinConv}} \\
\textbf{$N$} & \textbf{Pre.} & \textbf{Opt.} & \textbf{$\Delta$} [dB] & \textbf{Impr.} [\%] & \textbf{Pre.} & \textbf{Opt.} & \textbf{$\Delta$} [dB] & \textbf{Impr.} [\%] \\
\hline
10 & 12.6 & 11.6 & 1.0 &  8.1 &  8.1 & 7.7 & 0.4 &  4.4 \\
20 & 10.7 &  8.6 & 2.1 & 19.7 & 13.4 & 6.2 & 7.2 & 53.9 \\
30 &  9.4 &  6.7 & 2.7 & 28.3 &  8.2 & 6.8 & 1.4 & 17.2 \\
40 &  7.6 &  5.6 & 2.0 & 25.4 &  7.3 & 5.3 & 2.0 & 26.8 \\
50 &  5.6 &  4.4 & 1.2 & 21.7 &  8.7 & 8.2 & 0.5 &  5.9 \\
\hline
\end{tabular}
\end{table}

\section{Discussion}
    \label{sec:discussion}
    Overall, the results demonstrate that convergence-based refinement is most effective for selection strategies that focus on spatial coverage without explicitly accounting for the resulting \ac{REM} error. Strategies that already produce well-distributed measurement sets benefit only marginally, whereas weaker strategies improve substantially and reach performance levels comparable to more robust approaches.

The optimization gain depends on the relationship between $N$ and $k$: across all strategies, the average ratio $k/N$ lies between 0.75 and 0.94, with the optimizer exploiting redundancy within the candidate set. For \textit{MaxMinConv}, this ratio remains largely stable, while \textit{MinMax} shows improved performance for intermediate $N$ even as $k$ decreases, suggesting that increased spatial coverage enables smaller yet more informative subsets.

Although increasing $N$ generally improves performance, marginal gains diminish similarly across all strategies. While optimization can substantially enhance weaker pre-selection methods, it cannot fully compensate for inherent selection limitations, leaving a residual performance gap.

Rather than enumerating all subsets, Stage~2 combines an \ac{ILP}-based cardinality selection with a bounded local-search refinement. This heuristic remains tractable across the entire subset-size sweep, so the full sweep does not become a bottleneck. In our experiments, the local optimum it returns already yields the reported \ac{MAE} reductions. Stronger optimality could be pursued with metaheuristics (e.g., simulated annealing or genetic algorithms) without changing the two-stage structure.

The non-monotonic pre-selection performance of \textit{MaxMinConv} at $N=20$ and $N=50$ (Tab.~\ref{tab:transfer}) is structural rather than noise. The refinement optimizes a purely spatial criterion decoupled from the propagation field, so adding candidates need not reduce the reconstruction error. This redundancy is what Stage~2 exploits. Where sufficient degrees of freedom remain, the optimizer discards the uninformative points, as reflected by the large improvement at $N=20$, whereas the near-saturated set at $N=50$ leaves little room for correction. The optimization thus compensates for, rather than inherits, the pre-selection instability.

Tab.~\ref{tab:transfer} further reveals that the relative performance ranking between strategies can differ between simulation and real-world data. While \textit{MaxMinConv} consistently outperforms \textit{MinMax} in simulation (cf.\ Fig.~\ref{fig:opt_N}), \textit{MinMax} achieves comparable or lower \ac{MAE} on real-world data for $N \geq 30$. This inversion can be attributed to the mismatch between the simplified propagation model and the actual environment, which affects strategies differently depending on their spatial point configurations.

The evaluation is limited to a single rural open-field scenario. While the framework is conceptually scenario-agnostic, as only the propagation simulation model needs adaptation, urban and indoor environments introduce additional complexity (multipath, building shadowing, wall penetration) that may affect both pre-selection and optimization performance.

A methodological limitation concerns the reference used for the real-world evaluation. As no full-field ground truth is available, the reference \ac{REM} is itself interpolated from all accessible measurements using the same method employed for reconstruction. The reported error therefore partly reflects how closely a subset reproduces the full-set interpolation rather than the true field. This does not apply to the simulation-based results, for which the simulated field provides an independent reference. An unbiased assessment against held-out measurements is left for future work.

Two distinct sources of uncertainty affect the approach and impact it differently. Measurement noise enters only at \ac{REM} construction time and perturbs all candidate subsets in a comparable manner. Since the selection is driven by the simulated field, noise inflates the absolute \ac{MAE} but does not systematically bias which locations are selected. Model mismatch, by contrast, affects the selection itself, as it distorts the simulated field that guides the optimization. The gap between the simulated and real-world \ac{MAE} provides a measure of this mismatch, and the consistent relative improvement across this gap (Tab.~\ref{tab:transfer}) indicates that the selection is robust to the simplifications of the propagation model. More accurate models that capture additional effects such as vegetation shadowing would narrow this gap and further stabilize the per-$N$ performance.

The required inputs, namely terrain elevation (e.g., SRTM, Copernicus DEM) and antenna radiation patterns, are readily available, and capturing the relative spatial variation suffices to guide selection.

\section{Conclusion}
    \label{sec:conclusion}
    This work presents a two-stage measurement selection framework that improves \ac{REM} construction accuracy, quantified by \ac{MAE}, while reducing the number of required measurement locations. By combining heuristic pre-selection with simulation-based optimization, the framework exploits environmental information to identify high-quality measurement subsets from a reduced candidate set.

Both legacy selection strategies and the proposed convergence-based enhancements are effective pre-selection methods within the optimization framework. In particular, the enhanced strategies achieve earlier saturation of the \ac{REM} construction \ac{MAE}, indicating improved spatial coverage.

Evaluation using real-world measurement data confirms that configurations optimized in simulation are transferable to practical measurement scenarios. Across all evaluated strategies, the \ac{REM} construction \ac{MAE} is consistently reduced, demonstrating the practical viability of the proposed framework.

Future work will integrate explicit error constraints into the optimization, extend the evaluation to urban and indoor scenarios, empirically validate alternative interpolation methods within the framework, and investigate metaheuristic methods to further improve scalability beyond the current heuristic.

\section*{Acknowledgment}
This work was supported in part by the German Federal Ministry of Defence (BMVg) under project 5GOpportunity (grant no. 19OI22010B) and in part by the Bavarian Ministry of Economic Affairs, Regional Development and Energy (StMWi) under project EMSIC (grant no. DIK0517/03), and contributes to the 6G-Valley innovation cluster. The authors acknowledge the use of AI-assisted tools for language refinement. The authors alone are responsible for the content.

\bibliographystyle{IEEEtran}
\bibliography{references}

\end{document}